\documentclass[%
 aip,
 amsmath,amssymb,
 reprint,%
]{revtex4-1}

\usepackage{graphicx}
\usepackage{dcolumn}
\usepackage{bm}
\usepackage{amsmath}
\usepackage[utf8]{inputenc}
\usepackage[T1]{fontenc}
\usepackage{mathptmx}
\usepackage{textcomp}
\usepackage{gensymb}
\usepackage{tensor}

\begin{document}

\preprint{AIP/123-QED}

\title{A Paul Trap with Sectored Ring Electrodes for Experiments with Two-Dimensional Ion Crystals}

\author{M.K. Ivory}
\altaffiliation[Now at Sandia National Labs, Albuquerque, NM, 87185, USA]{}
\author{A. Kato}
 \email{kato@uw.edu}
\affiliation{University of Washington Department of Physics, Seattle , Washington, USA, 98115}
\author{A. Hasanzadeh}
\affiliation{University of Washington Department of Physics, Seattle , Washington, USA, 98115}
\author{B. Blinov}
\affiliation{University of Washington Department of Physics, Seattle , Washington, USA, 98115}

\date{\today}

\begin{abstract}
We have developed a trapped ion system for producing two-dimensional (2D) ion crystals for applications in scalable quantum computing, quantum simulations, and 2D crystal phase transition and defect studies.  The trap is a modification of a Paul trap with its ring electrode flattened and split into eight identical sectors, and its two endcap electrodes shaped as truncated hollow cones for laser and imaging optics access. All ten trap electrodes can be independently DC-biased to create various aspect ratio trap geometries. We trap and Doppler cool 2D crystals of up to 30 Ba${^+}$ ions and demonstrate the tunability of the trapping potential both in the plane of the crystal and in the transverse direction.

\end{abstract}

\maketitle

Trapped ion qubits are one of the leading technologies for scalable quantum computing and quantum simulations due to their extremely long coherence times and high-fidelity state initialization, control, and readout. The linear ion trap is the workhorse of trapped ion quantum computing, and has led to the successful demonstration on entanglement of up to 20 qubits \cite{Friis2018}, universal quantum computation \cite{Hanneke2010}, and quantum simulations \cite{Richerme2014}.
Proposed architectures exist that promise scaling up the linear ion trap system to hundreds or even thousands of qubits, such as the modular MUSIQC architecture\cite{Monroe2014}  and the QCCD architecture with ion shuttling\cite{Kielpinski2002}.

\begin{figure}[hb!]
\includegraphics[width=81mm]{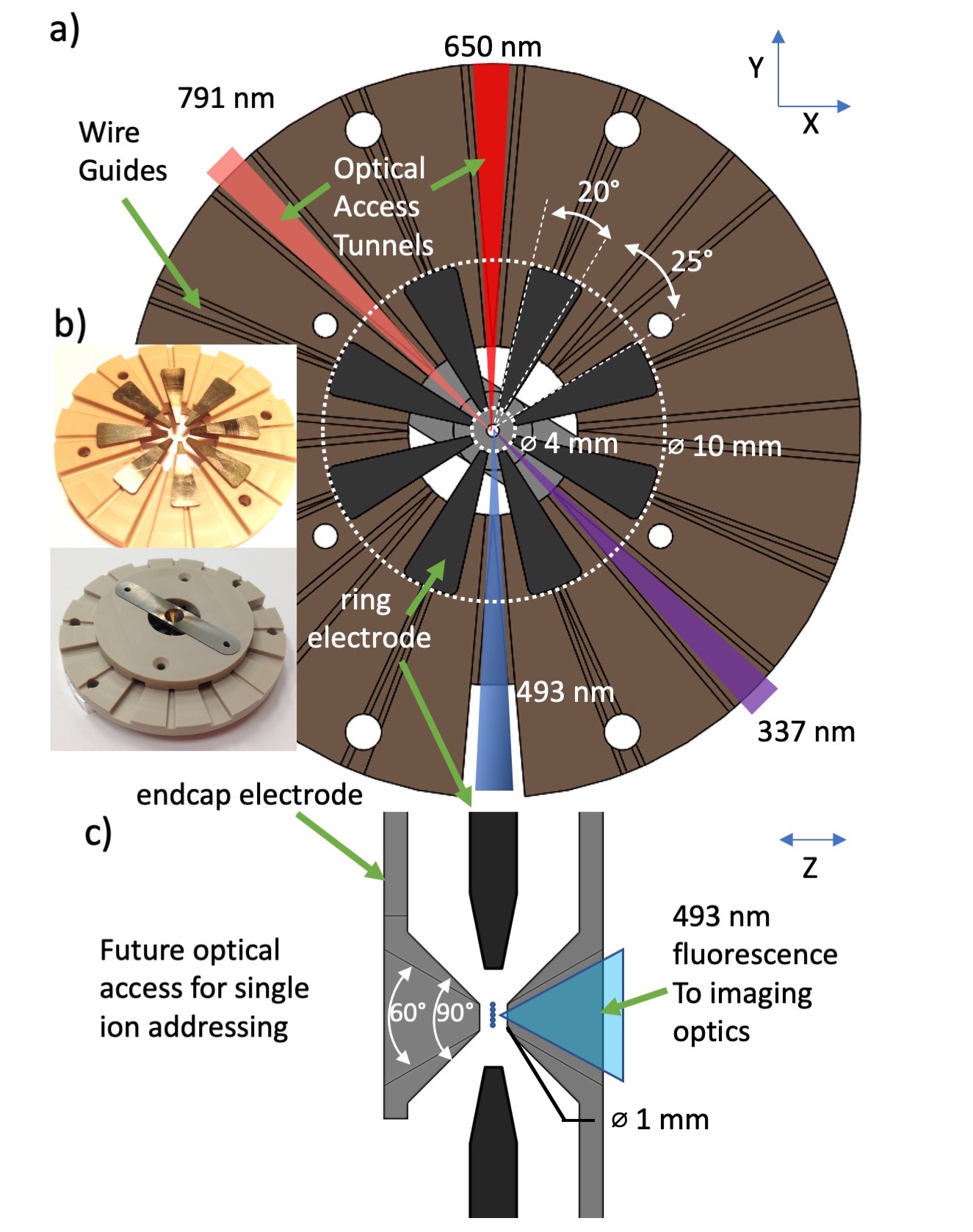}
\caption{\label{fig:Figure_1} Design and schematic layout of the trap. (a) Planar (XY) cross section of the trap assembly. The eight ring electrode sectors are shown in dark gray, and the PEEK holder in brown. In between each ring segment, optical access tunnels allow for lasers to pass through. The ring sectors are spot welded to wires, which are fed through wire guides. The 791 nm and 337 nm laser beams are used for photoionization of barium, while 493 nm and 650 nm are used for Doppler cooling the ions. (b) Photographs of the trap housed in the PEEK structure. (c) Cross section of the trap in XZ plane. Ions are imaged through one endcap, while the other is reserved for individual ion addressing.}
\end{figure}
 
2D trapped ion crystals offer a higher qubit density, and have been proposed as a way to scale to over 100 qubits  \cite{Porras2006}. Previously, such geometries were avoided since there is only a single point, the RF null, where micromotion can be minimized. Ions trapped away from the RF null inevitably experience excess micromotion that leads to  the decoherence of qubits and limits quantum gate fidelity \cite{Yoshimura2015}. However, for trap geometries with transverse symmetry, a plane exists where there is a node in the transverse electric field. Ion crystals in this plane only experience excess planar micromotion, while excess transverse micromotion may be minimized. Therefore, lasers that propagate along the transverse axis are not significantly Doppler-shifted in the ion's frame.  Unavoidable in-plane excess micromotion causes an intensity modulation of the laser light that interacts with the ion, but recent work outlines a method to compensate for this by employing a series of segmented laser pulses\cite{Wang2015}. This method is predicted to achieve two-qubit gates with >99.99$\%$ fidelity even between ions at the edge of a 127-ion crystal. Applying segmented laser pulse techniques has also recently achieved the fastest two-qubit gates in one-dimensional trapped ion systems to date \cite{Shafer2015}.

\begin{figure*}[ht]
\includegraphics[width=\textwidth]{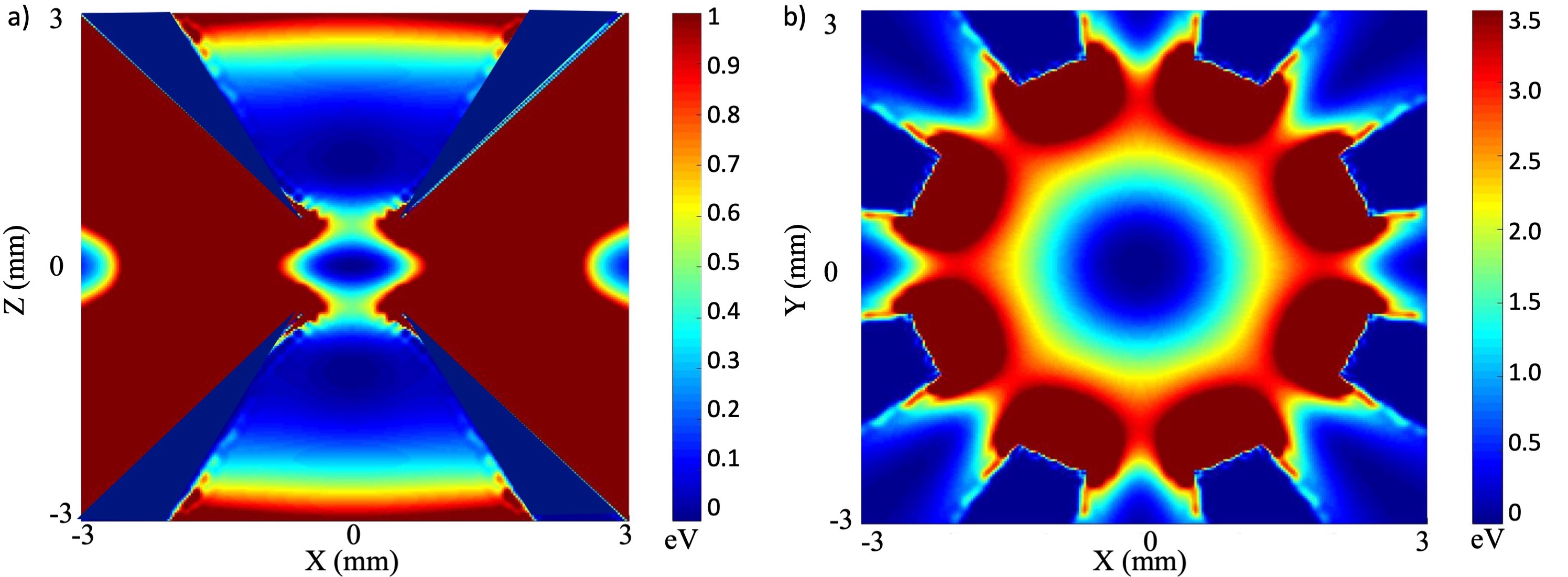}
\caption{\label{fig:Figure_2} Simulated RF pseudopotential of the trap with 1000 V of RF at 12.47 MHz. (a) XZ plane (b) XY plane. The pseudopotential is symmetric about the Z axis and under 45\degree rotations in the XY plane. For these RF voltages and no DC confinement, we obtain $\alpha$=2.3.  We note that a DC potential is necessary to increase the aspect ratio.}
\end{figure*}

2D trapped ion crystals have been previously studied  in linear ion traps in the plane perpendicular to the trap axis, where it has been reported that 19 ions have been trapped in a 2D formation \cite{Block2000}, and indirectly observed that 34 have been trapped \cite{Okada2010}. A challenge associated with linear traps is that the imaging plane is not parallel to the plane of the crystal due to the presence of the endcaps. It has also been reported that $\sim$150 ions have been trapped in a 2D formation in a microfabricated trap \cite{Szymanski2012}. 

Here, we present an experimental demonstration of a system for creating 2D ion crystals as the first step toward scalable quantum computing experiments.  Our approach addresses two main challenges:  achieving 1) the necessary aspect ratios for transverse-to-planar confinement,  and 2) the transverse optical access for single ion addressing and imaging. We do so by employing a modified ring trap similar to that described in a detailed study of 2D crystals for quantum simulations\cite{Yoshimura2015}, using  bored endcaps to allow for transverse optical access and imaging as well as strong transverse confinement.  Our trap design is shown in Figure 1.

Our trap differs notably in several ways. The ring electrode is divided into eight sectors while each of the endcaps are composed of a single piece of material. Independent DC bias voltages may be applied to each of the 10 electrodes. In this way we can independently tune transverse and in-plane trapping potentials. The sectors also permit in-plane optical access.  The trap ring sectors are 20{\degree}  metal wedges with 25{\degree}  spacing, with an ID of 4 mm and OD of 10 mm. The endcaps are truncated hollow cones with a numerical aperture of 0.5 and separation distance of 1 mm for optical access  and imaging from either side of the plane of the crystal. 

In order to achieve 2D trapped ion crystals, we estimate the transverse (Z)  to planar (XY-plane) secular frequency ratio $\alpha=\omega^{}_{z}/\omega^{}_{r}$ using

\begin{equation}
\alpha^{2} = \sqrt{\frac{96N}{\pi^{3}\omega^{3}_{1}}},
\end{equation}

\noindent where N is the number of ions and $\omega^{}_{1} = 1.11$ is a parameter below which  transitions to 2D formations occur\cite{Dubin1993}. While this equation is more accurate for large ion numbers, it serves as a guideline of the aspect ratio necessary to produce 2D crystals for smaller ion numbers\cite{Okada2010}. We target $\alpha>3$ to get 2D crystals of $\sim$30 ions, according to equation 1. The numerical simulation of the trap’s RF pseudopotential with $\alpha=2.3$ is shown in Figure 2. This aspect ratio is not high enough to produce large planar crystals, and hence the application of a DC potential to the endcaps is necessary.

In addition to positioning the crystal in the center of the trap, the ring segments may be used to change the in-plane potential shape of the 2D crystal. The potential can be made radially symmetric to make crystals with concentric rings, or be compressed to force ions towards a line in the XY plane. As ions are compressed towards a linear shape in the XY plane, one in-plane trap secular frequency is increased while the second is decreased. As a result $\alpha$ may fall below the
necessary ratio and the crystal may begin to zig-zag in and out-of plane.
Ion-ion spacing can be tuned by adjusting the overall trap strength with the RF voltage, by adjusting $\alpha$ with the DC voltages, or by changing the number of ions in the crystal. We target spacings in the range  6.5-11 $\mu$m for single ion addressing. Spacing is limited on the lower end due to the constraint that crystals must remain planar. 

 \begin{figure*}[ht]
\includegraphics[width=\textwidth]{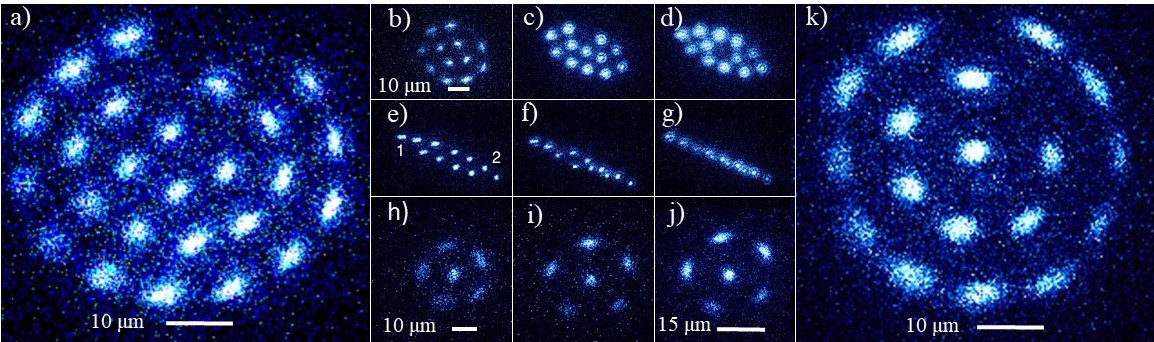}
\caption{\label{fig:Figure_3} Trapped ion crystals produced at different trapping potentials.  (a),(b),(h) begin with $\sim$1000 V of RF at 12.47 MHz.  (a) A 29-ion crystal. All ring sector  electrodes are at -152 V with endcaps at -145 V and -150 V, a configuration found to compensate stray DC electric fields and allow crystallization near the center of the trap. (b-g) 13 ion crystal with a voltage increase on two opposing sector electrodes, changing the shape of the potential and crystal shape. In (f) and (g) the ions begin to zig-zag out of plane. Two opposing sector electrodes are each increased by 15 V in 3V increments. (h-j) 6 ion crystal with increasing voltage on each endcap electrode by 5V in 2.5V increments. (k) 20 ion crystal with  reduced asymmetry.  Non-fluorescing ions are other isotopes of Barium that are not addressed by the cooling laser. Image scales are consistent in frames (b-g) and in (h-j).
}
\end{figure*}

The trap components were built by the University of Washington Physics Department Instrument Shop. The electrodes were Electric Discharge Machined (EDM) out of stainless steel and electropolished smooth to avoid charge buildup at burrs or edges.  Kapton-insulated wires were spot welded to connect each electrode to a 10-pin vacuum feedthrough. The electrodes are housed in an a machined structure composed of polyetheretherketone (PEEK) that contains radial tunnels for wire guides, laser access, and atomic beam for ion loading. The trap assembly is mounted in a 4.5” spherical octagon vacuum chamber (Kimball Physics part no. MCF450-SphOct-E2A8).

The system for producing, ionizing, and cooling Ba${^+}$ ions is similar to\cite{Graham2014}, with the notable exception that 493 nm  light for Doppler cooling is produced directly via an External Cavity Diode Laser (ECDL) using a Sharp GH04850B2G diode. The fluorescence of ions is imaged using an Andor iXon Electron-Multiplying Charge-Coupled Device (EMCCD) camera. We use a Mitutoyo 0.28 NA long working distance microscope objective lens in conjunction with a 25 mm doublet lens to image the ions, and we measure a magnification of 78x. The RF voltage at 12.47 MHz is applied to the two endcap electrodes via a helical resonator with a Q factor of $\sim$225.  The helical resonator contains 2 interwound coils that allow for each endcap electrode to be DC biased independently. The RF voltage can be raised up above $\sim$1000 V, where it is eventually limited by arcing in the vacuum feedthrough. Each of the 8 ring electrodes are  grounded for RF frequencies using RC lowpass filters with cutoff frequencies of 3 MHz.

${^{138}}$Ba${^+}$ ions are loaded into the trap at a rate of approximately 1 ion/s using two-step resonant photoionization with a 791 nm ECDL and a 337 nm nitrogen laser. We observed 2D crystal geometries that  agree with  predicted shell structure \cite{Buluta2009,Bedanov1994}. A selection of these crystals is shown in Figure 3. For example, in Figure 3 (a), a 29-ion crystal is shown. For the trapping potential used to produce such crystal we measure single ion planar secular frequencies of 203 kHz and 221 kHz through application of a tickle voltage delivered to one of the ring sectors. From equation 1, we infer that the transverse trap frequency is $\sim$600 kHz with the parameters given in Figure 2 (a). Laser-cooled crystal lifetimes of several hours have been observed, with background collisions occasionally (approximately once every 10 minutes) leading to dark ions appearing in the crystal structure or reordering of ions.

In Figure 3 we also demonstrate the tunability of our trap aspect ratios. In panels (b) through (g), we increase the voltage on two opposing segments of the ring by a total of 15 V in steps of 3V. The shape of the ion crystal changes from a nearly circular in (b) to a linear crystal with in- and out-of-plane zig-zag in (f) and (g). The effect of increasing the endcap bias voltage by 5 V in steps of 2.5 V is shown in Figure 3 (h-j). The ion spacing in the plane of the crystal increases as the endcap voltage (radial confinement) is increased (decreased). In Figure 3(k) we show a 20 ion crystal with nearly equal planar trapping frequencies.

The amplitude of the excess in-plane micromotion is given by $q_r d/2$, where $q_r$ is the radial Matthieu parameter, and d is the distance from the ion to the trap center\cite{Chou2017}. Based on the trap voltages, we estimate the scale of the planar micromotion to be 0.051 $\mu$m for every 1 $\mu$m of displacement from the trap center. In Figure 3(e), the distance between ions labelled 1 and 2 is approximately 54 $\mu$m. This yields an estimate of a $\sim$2.8 $\mu$m of additional excess micromotion between these ions, which is in very good agreement with a measured increase in ion image size of about 2.6 $\mu$m (1/$e^2$ brightness) between ions 1 and 2.

In summary, we outline the design and construction of a trap for producing 2D ion crystals, and demonstrate the capability of the trap to produce such crystals with broadly tunable parameters. Future work will include characterizing micromotion in the trap and demonstrating single ion addressing for scalable quantum logic operations.
	
This work was supported by the National Science Foundation Grant No. 1505326, University of Washington Royalty Research Fund, and the Mistletoe Fellowship. The authors  would like to thank Liudmila Zhukas, Jennifer Lilieholm, and Gabriel Moreau for useful insights. 

\bibliography{References}

\end{document}